\documentclass[12pt]{article}

\usepackage{amsmath,amssymb,theorem}

\theoremstyle{plain}

{\theorembodyfont{\rmfamily}

\input amssym.def   
\input amssym       

\def\qed{\hfill\ \vbox{\hrule\hbox{\vrule\kern4pt\vbox{\kern4pt{}
\kern4pt}\kern4pt\vrule}\hrule}}

\addtolength{\oddsidemargin}{-1.0cm}
\addtolength{\textwidth}{+2cm}

\def\be{\begin{equation}}
\def\ee{\end{equation}}
\def\bc{\begin{center}}
\def\ec{\end{center}}
\def\nd{\noindent}

\def\R{{\Bbb R}}

\def\C{{\Bbb C}}
\def\oH{\buildrel\circ\over H}

\begin{document}

J.Math. Phys., 40, N8, (1999), 3876-3880

\title{ Inverse problem for an inhomogeneous \\
Schr\"odinger equation
\thanks{key words: inverse problem, inverse scattering, Schr\"odinger
 equation  }
\thanks{PACS numbers 02.30.Hq, 03.65.Nk }
}

\author{A.G. Ramm\\
 Mathematics Department, Kansas State University, \\
 Manhattan, KS 66506-2602, USA\\
ramm@math.ksu.edu\\
}

\date{}

\maketitle\thispagestyle{empty}

\begin{abstract}
Let $(\ell-k^2) u=-u''+q(x)u-k^2u=\delta(x)$, $x\in \R$,
$\frac{\partial u}{\partial|x|}-iku\to 0$, $|x|\to\infty$.

Assume that the potential $q(x)$ is real-valued and compactly supported:
$q(x)=\overline{q(x)}$, $q(x)=0$ for $|x|\geq 1$,
$\int^1_{-1}|q|dx<\infty$, and that $q(x)$ produces no
bound states.
Let $u(-1,k)$ and $u(1,k)\quad\forall k>0$ be the data.

\vspace{.15in}
\noindent{\bf Theorem.}{\it Under the above assumptions
these data determine $q(x)$ uniquely.}

\end{abstract}


\section{Introduction}

For several decades the following inverse problems of
practical interest are open. Let
$$ \nabla^2u+k^2u+k^2v(x)u=-\delta(x)\ \hbox{in}\  \R^3,
   \eqno(1.1) $$
$u$ satisfies the radiation condition at infinity, $v(x)$ is a
compactly supported piecewise-smooth function,
$\hbox{supp}v\subset\R^3_-:=\{x:x_3<0\}$.

The data are the values $u(x_1,x_2,0,k)$ for all
$\widehat x:=(x_1,x_2)\in \R^2$ and $k>0$.

(IP1) The inverse problem is: 

{\it Given the data, find $v(x)$.}

Uniqueness of the solution to this problem is not proved.
IP1 is not overdetermined: the data is a function of three variables,
and $v(x)$ also is.

A similar inverse problem can be formulated:
Let
$$ \nabla^2u+k^2u-q(x)u=0\ \hbox{in}\ \R^3,
  \eqno(1.2) $$
$$ u=e^{ik\alpha\cdot x} + A(\alpha',\alpha,k)
   \frac{e^{ikr}}{r}+o \left(\frac{1}{r}\right),
   r:=|x|\to\infty, \alpha'=\frac{x}{r},
   \eqno(1.3) $$
where $\alpha\in S^2$ is a given unit vector, $q(x)$ is a real-valued
piecewise-smooth function, \linebreak
$\hbox{supp}\, q(x)\subset B_a:=\{x:|x|\leq a\},S^2$
is the unit sphere.

(IP2) {\it Given $A(\alpha',\alpha_0,k)$ for all $\alpha'\in S^2$,
all $k>0$ and a fixed $\alpha=\alpha_0\in S^2$, find $q(x)$.}

The uniqueness of the solution to (IP2) is not proved.

The third problem is:

Let
$$ \nabla^2u+k^2u-q(x)u=-\delta(x)
   \ \hbox{in}\ \R^3,
   \eqno(1.4) $$
$u$ satisfies the radiation condition, $q(x)$ is the same as in (IP2).

The data are the values $u(x,k)\bigg|_{|x|=a}$.

(IP3) {\it Given the data $u(x,k)\bigg|_{|x|=a}$ for all
$k>0$ and all $x$ on the sphere $S_a:=\{x:|x|=a\}$,
find $q(x)$.}

Uniqueness of the solution to (IP3) is not proved.

An overview of inverse problems and references one can find in 
[1]- [3].

The purpose of this paper is to study the one-dimensional analog
of (IP3) and to prove for this analog a uniqueness theorem. The 
one-dimensional analog of (IP3) corresponds to a plasma equation in a layer.

Let
$$ \ell u-k^2u:=-u''+q(x)u-k^2u=\delta(x),\quad x\in\R^1,
   \eqno(1.5) $$

$$ \frac{\partial u}{\partial|x|} -iku\to 0,\quad |x|\to\infty.
   \eqno(1.6) $$

Assume that $q(x)$ is a real-valued function,
$$ q(x)=0\ \hbox{for}\ |x|>1,\quad q\in L^\infty[-1,1].
   \eqno(1.7) $$

Suppose that the data
$$ \{u(-1,k), u(1,k)\},\quad \forall k>0
   \eqno(1.8) $$
are given.

The inverse problem analogous to (IP3) is:

(IP) Given the data (1.8), find $q(x)$.

This problem, as well as (IP1)-(IP3), is of practical interest.
One can think about finding the properties of an
inhomogeneous slab (the governing equation is plasma equation)
from the boundary measurements of the field,
generated by a point source inside the slab.

In the literature there are many results concerning
various inverse problems for the homogeneous version of equation (1.5),
but it seems that no results concerning (IP) are known.

Assume that the self-adjoint operator
$\ell=-\frac{d^2}{dx^2}+q(x)$ in $L^2(\R)$
has no negative eigenvalues (this is the case
when $q(x) \geq 0$, for example).
The operator $\ell$ is the closure in $L^2(\R)$
of the symmetric operator $\ell_0$ defined on $C^\infty_0(\R^1)$
by the formula $\ell_0u=-u''+q(x)u$.
Our result is:

\vspace{.15in}
\nd{\bf Theorem 1.}
{\it
Under the above assumptions IP has at most one solution.
} 

\vspace{.15in}
\nd{\bf Proof of Theorem 1:}
The solution to (1.5)-(1.6) is
\begin{equation}
   u=\left\{
     \begin{split}
       \frac{g(k)}{[f,g]} f(x,k),\quad x>0, \\
       \frac{f(k)}{[f,g]} g(x,k),\quad x<0.
     \end{split}
  \right. \tag{2.1} \end{equation}
Here $f(x,k)$ and $g(x,k)$ solve homogeneous version
of equation (1.5) and
have the following asymptotics:
$$ f(x,k)\sim e^{ikx},\quad x\to +\infty,\quad g(x,k)\sim e^{-ikx},
   \quad x\to -\infty,
   \eqno(2.2) $$
$$ f(k):=f(0,k), \quad g(k):=g(0,k),
   \eqno(2.3) $$
$$ [f,g]:=fg'-f'g=-2ika(k),
   \eqno(2.4) $$
where the prime denotes differentiation with
respect to $x$-variable,
and $a(k)$ is defined by the equation
$$ f(x,k)=b(k)g(x,k)+a(k)g(x,-k).
   \eqno(2.5) $$
It is known (see for example [4]) that
$$ g(x,k)=-b(-k)f(x,k)+a(k)f(x,-k),
   \eqno(2.6) $$
$$ a(-k)=\overline{a(k)},\quad b(-k)=\overline{b(k)},
   \quad |a(k)|^2=1+|b(k)|^2, \quad k\in\R,
   \eqno(2.7) $$
$$ a(k)=1+O(\frac 1k),\quad k\to\infty,\quad k\in\C_+;
   \quad b(k)=O(\frac 1k),\quad |k|\to\infty,\quad k\in\R,
   \eqno(2.7')$$
$$ [f(x,k),g(x,-k)]=2ikb(k),\quad [f(x,k),g(x,k)]=-2ika(k),
   \eqno(2.8) $$
$a(k)$ in analytic in $\C_+$, $b(k)$ in general does not admit analytic
continuation from $\R$, but if $q(x)$ is compactly supported,
then $a(k)$ and $b(k)$ are analytic functions of $k\in\C\setminus 0$.

The functions
$$ A_1(k):=\frac{g(k)f(1,k)}{-2ika(k)},
   \quad A_2(k):=\frac{f(k) g(-1,k)}{-2ika(k)}
   \eqno(2.9) $$
are the data,
they are known for all $k>0$. Therefore one can assume the functions
$$ h_1(k):=\frac{g(k)}{a(k)},\quad h_2(k):=\frac{f(k)}{a(k)}
   \eqno(2.10) $$
to be known for all $k>0$ because
$$ f(1,k)=e^{ik},\quad g(-1,k)=e^{ik},
   \eqno(2.11) $$
as follows from the assumption (1.7) and from (2.2).

From (2.10), (2.6) and (2.5) it follows that
$$ a(k) h_1(k)=-b(-k)f(k)+a(k)f(-k)
   =-b(-k)h_2(k)a(k) +h_2(-k)a(-k)a(k),
   \eqno(2.12) $$
$$ a(k)h_2(k) =b(k)a(k)h_1(k) +a(k)h_1(-k)a(-k).
   \eqno(2.13) $$
From (2.12) and (2.13) it follows:
$$ -b(-k)h_2(k) +h_2(-k)a(-k) =h_1(k),
   \eqno(2.14) $$
$$ b(k)h_1(k) +a(-k)h_1(-k) =h_2(k).
   \eqno(2.15) $$
Eliminating $b(-k)$ from (2.14) and (2.15), one gets:
$$ a(k)h_1(k)h_2(k) +a(-k)h_1(-k)h_2(-k)
   =h_1(k)h_1(-k) +h_2(-k)h_2(k),
   \eqno(2.16) $$
or
$$ a(k)=m(k)a(-k) +n(k),\quad k\in\R
   \eqno(2.17) $$
where
$$ m(k):=-\frac{h_1(-k)h_2(-k)}{h_1(k)h_2(k)},
   \quad n(k):=\frac{h_1(-k)}{h_2(k)} + \frac{h_2(-k)}{h_1(k)}.
   \eqno(2.18) $$

Problem (2.17) is a Riemann problem (see [5] for the theory
of this problem )
 for the pair $\{a(k),a(-k)\}$,
the function $a(k)$ is analytic in $\C_+:=\{k:k\in\C,Imk>0\}$ and
$a(-k)$ is analytic in $\C_-$. The functions
$a(k)$ and $a(-k)$ tend to one as $k$ tends to infinity
in $\C_+$ and, respectively, in $\C_-$, see equation (2.7$^\prime$).

The function $a(k)$ has finitely many simple zeros at the points
$ik_j,1\leq j\leq J$, $k_j>0$, where $-k^2_j$ are the negative
eigenvalues of the operator $\ell$ defined by the differential
expression $\ell u=-u''+q(x)u$ in $L^2(\R)$.

The zeros $ik_j$ are the only zeros of $a(k)$ in the upper half-plane $k$.

Define
$$ ind\, a(k):=\frac{1}{2\pi i}\int^\infty_{-\infty} d\,\ln\,a(k).
   \eqno(2.19) $$
One has
$$ ind\, a=J,
   \eqno(2.20) $$
where $J$ is the number of negative eigenvalues of the
operator $\ell$, and, using (2.10), (2.20) and (2.18), one gets
$$ ind\,m(k)=-2[ind\,h_1(k)+ind\, h_2(k)]
   =-2[ind\,g(k)+ind\, f(k)-2J].
   \eqno(2.21) $$
Since $\ell$ has no negative eigenvalues, it follows that $J=0$.

In this case $ind\,f(k)=ind\,g(k)=0$ (see Lemma 1 below),
so $ind\,m(k)=0$, and $a(k)$ is uniquely recovered from the data
as the solution of (2.17) which tends to 
one at infinity, see equation (2.7$^\prime$). If $a(k)$ is found,
then $b(k)$ is uniquely determined by equation (2.15) and so
the reflection coefficient $r(k):=\frac{b(k)}{a(k)}$
is found. The reflection coefficient determines a compactly
supported $q(x)$ uniquely \cite{R1}.

To make this paper self-contained, let us outline a proof
of the last claim using an argument different from the one
given in [2]. 

If $q(x)$ is compactly supported, then the reflection coefficient
$r(k):=\frac {b(k)}{a(k)}$ is meromorphic. Therefore, its values
for all $k>0$ determine uniquely $r(k)$ in the whole
complex $k$-plane as a meromorphic function. The poles
of this function in the upper half-plane are the numbers
$ik_j, j=1,2,...,J$. They determine uniquely the numbers $k_j, 1\leq j
\leq J,$ which are a part of the standard scattering data
$\{r(k), k_j, s_j, 1\leq j \leq J\}$, where $s_j$ are the norming
constants.

Note that if $a(ik_j)=0$ then $b(ik_j)\neq 0$: otherwise
equation (2.5) would imply $f(x,ik_j)\equiv 0$ in contradiction
to the first relation (2.2).

 If $r(k)$ is meromorphic, then the norming constants can
be calculated by the formula $s_j=-i\frac {b(ik_j)}{\dot a(ik_j)}=
-i Res_{k=ik_j} r(k)$, where the dot denotes differentiation
with respect to $k$, and $Res$ denotes the residue. So,
for compactly supported potential the values of $r(k)$ for
all $k>0$ determine uniquely the standard scattering data,
that is, the reflection coefficient, the bound states $-k_j^2$
and the norming constants $s_j$, $1\leq j \leq J.$
These data determine the potential uniquely.	

Theorem 1 is proved. \qed

\nd{\bf Lemma 1.}
{\it
If $J=0$ then $ind\,f=ind\,g=0$.
}

\nd{\bf Proof.}
We prove $ind\,f=0$. The proof of the equation $ind\,g=0$ is similar.
Since $ind\,f(k)$ equals to the number of zeros of $f(k)$ in $\C_+$,
we have to prove that $f(k)$ does not vanish in $\C_+$.
If $f(z)=0$, $z\in\C_+$, then 
$z=ik$, $k>0$, and $-k^2$ is an eigenvalue of the
operator $\ell$ in $L^2(0,\infty)$ with the boundary condition
$u(0)=0$.

From the variational principle one
can find the negative eigenvalues of the operator $\ell$
in $L^2(\R_+)$ with the Dirichlet condition at $x=0$ as consequitive
minima of the quadratic functional. The minimal eigenvalue is:
$$ -k^2=inf\int^\infty_0
   \left[ u^{\prime 2}+q(x)u^2 \right] dx:=
   \kappa_0,
   \quad u\in \oH\kern-.02in{}^1(\R_+),\quad ||u||_{L^2(\R_+)}=1,
   \eqno(2.22) $$
where $\oH\kern-.02in{}^1(\R_+)$ is the Sobolev space of
$H^1(\R_+)$-functions
satisfying the condition $u(0)=0$.

On the other hand, if $J=0$, then
$$ 0\leq inf\,\int^\infty_{-\infty} [u^{\prime 2}+q(x)u^2]\,dx:=
\kappa_1,
   \quad u\in H^1(\R), \quad ||u||_{L^2(\R)}=1.
   \eqno(2.23) $$
Since any element $u$ of $\oH\kern-.02in{}^1(\R_+)$ can be considered as
an element of $H^1(\R)$ if one extends $u$ to the whole axis by setting
$u=0$ for $x<0$, it follows from the variational definitions (2.22)
and (2.23) that
$\kappa_1\leq \kappa_0$. Therefore, if  $J=0$, 
then $\kappa_1\geq 0$ and therefore $\kappa_0 \geq 0$.
This means that operator $\ell$ on $L^2(\R_+)$ with the Dirichlet condition
at $x=0$ has no negative eigenvalues.
This means that $f(k)$ does not have zeros in $\C_+$, if $J=0$. 
Thus $J=0$ implies $ind\,f(k)=0$.

Lemma 1 is proved. \qed

\nd{\bf Remark 2.}
The above argument shows that in general
$$ ind\, f\leq J \quad\hbox{and}\quad ind\,g\leq J,
   \eqno(2.24) $$
so that (2.21) implies
$$ ind\,m(k)\geq 0.
   \eqno(2.25) $$
Therefore the Riemann problem (2.17) is always solvable.

\end{document}